\RequirePackage{fix-cm}
\documentclass[smallextended]{svjour3}       
\smartqed  
\usepackage{graphicx}

 \journalname{Hyperfine Interactions}
\begin{document}

\title{GAPS, low-energy antimatter for indirect dark-matter search
}

\author{
E.~Vannuccini$^{1}$ 
\and T.~Aramaki$^{2}$  \and  R.~Bird$^{3}$  \and M.~Boezio$^{4}$ \and 
S.E.~Boggs$^{5}$  \and V.~Bonvicini$^{4}$  \and D.~Campana$^{6}$    \and W.W.~Craig$^{8}$  \and  
P.~von~Doetinchem$^{9}$  \and E.~Everson$^{3}$ \and  L.~Fabris$^{10}$  \and  F.~Gahbauer$^{11}$ \and C.~Gerrity$^{9}$  \and
H.~Fuke$^{12}$  \and C.J.~Hailey$^{11}$ \and T.~Hayashi$^{3}$ \and C.~Kato$^{13}$  \and A.~Kawachi$^{14}$ \and  M.~Kozai$^{12}$  \and A.~Lowell$^{5}$  \and 
M.~Martucci$^{15}$  \and  S.I.~Mognet$^{16}$  \and R.~Munini$^{4}$  \and  K.~Munakata$^{13}$  \and S.~Okazaki$^{12}$  \and R.A.~Ong$^{3}$ \and  G.~Osteria$^{6}$  \and K.~Perez$^{7}$  \and S.~Quinn$^{3}$ \and J.~Ryan$^{3}$ \and  V.~Re$^{17}$  \and F.~Rogers$^{7}$  \and N.~Saffold$^{11}$  \and Y.~Shimizu$^{18}$  \and R.~Sparvoli$^{15}$ \and A.~Stoessl$^{9}$  \and  A.~Yoshida$^{19}$ \and T.~Yoshida$^{12}$  \and G.~Zampa$^{4}$  \and J.~Zweerink$^{3}$
          }

\institute{
E. Vannuccini \\
\email{vannuccini.elena@fi.infn.it} \\
\at $^{1}$ INFN Firenze,  Edificio di Fisica Sperimentale, Via Bruno Rossi 3, 50019 Sesto Fiorentino, Italia\\
\at  $^{2}$ Stanford Linear Accelerator Center,  2575 Sand Hill Road,  Menlo Park, CA 94025, USA\\
\at $^{3}$ UCLA Physics and Astronomy,  430 Portola Plaza,  University of California,  Los Angeles, CA 90095-1547,  USA \\
\at $^{4}$INFN Trieste,  AREA di Ricerca di Trieste,  Padriciano 99,  I-34149 Trieste,  Italy\\
\at $^{5}$ Department of Physics,  MH 3561,  University of California, San Diego, 9500 Gilman Drive, La Jolla, CA 92093-0354, USA\\
\at $^{6}$ INFN Napoli, Complesso Universitatrio di M.S. Angelo, Ed. 6, Via Cintia, 80126 Napoli, Italy\\
\at $^{7}$ MIT  Laboratory of Nuclear Science, 77 Massachusetts Ave, Building 26, Cambridge, MA 02139,  USA\\
\at $^{8}$ University of California at Berkeley, Space Sciences Laboratory, 7 Gauss Way, Berkeley, CA 94720-7450, USA\\
\at $^{9}$ Department of Physics and Astronomy, 2505 Correa Rd, University of Hawaii at Manoa, Honolulu, HI 96822, USA\\
\at $^{10}$  Oak Ridge National Laboratory, Nuclear Isotope and Isotope Technology Division, 1 Bethel Valley Road MS-6010, Oak Ridge, TN 37831, USA\\
\at $^{11}$ Columbia University Astrophysics Lab, 550 West 120th street, Columbia University, New York, NY 10027, USA\\
\at $^{12}$Institute of Space and Aeronautical Science, Japan Aerospace Exploration Agency, 3-1-1 
nodai, Chuo-ku, Sagamihara, Kanagawa 252-5210, Japan\\
\at $^{13}$Shinshu University, 3-1-1, Asahi, Matsumoto, Nagano 390-8621, Japan\\
\at $^{14}$Tokai University, 4-1-1 Kitakaname, Hiratsuka, Kanagawa 259-1292, Japan\\
\at  $^{15}$University of Roma ``Tor Vergata'' and INFN Roma ``Tor Vergata'', Rome, Italy\\
\at  $^{16}$ Pennsylvania State University, USA\\
\at $^{17}$ University of Bergamo, Department of Engineering and Applied Sciences, viale Marconi 5, 24044 Dalmine, Bergamo,  Italy\\
\at $^{18}$ Kanagawa University, 3-27-1 Rokkakubashi, Kanagawa-ku, Yokohama, Kanagawa 221-8686, Japan\\
\at $^{19}$ Aoyama Gakuin University 5-10-1 Fuchinobe, Chuo-ku, Sagamihara, Kanagawa 252-5258, Japan\\
}

\date{Received: date / Accepted: date}

\maketitle

\begin{abstract}
The General Antiparticle Spectrometer (GAPS) is designed to carry out indirect dark matter search by measuring low-energy cosmic-ray antiparticles.
Below a few GeVs the flux of antiparticles produced by cosmic-ray collisions with the interstellar medium
is expected to be very low and several well-motivated beyond-standard models predict a sizable contribution to the antideuteron flux.
GAPS is planned to fly on a long-duration balloon over Antarctica in the austral summer of 2020.
The primary detector is a $\sim$1m$^3$ central volume containing planes of Si(Li) detectors. This volume is surrounded by a time-of-flight system to both trigger the Si(Li) detector and reconstruct the particle tracks.
The detection principle of the experiment relies on the identification of the antiparticle  annihilation pattern.
Low energy antiparticles slow down in the apparatus and they are captured in the medium to form exotic excited atoms, which de-excite by emitting characteristic X-rays. 
Afterwards they undergo nuclear annihilation, resulting in a star of pions and  protons. 
The simultaneous measurement of the stopping depth and the dE/dx loss of the primary antiparticle, of the X-ray energies and of the star particle-multiplicity provides very high rejection power, that is critical in rare-event search.
GAPS will be able to perform a precise measurement of the cosmic antiproton flux below 250~MeV, as well as a sensitive search for antideuterons.
\keywords{Dark Matter \and Cosmic Rays \and Antiparticle \and Antideuteron \and Antiproton \and Exotic Atoms \and GAPS \and Balloon Experiment}
\end{abstract}


\section{Introduction}
\label{intro}

According to many consistent astrophysical and cosmological observations, about one fourth  of our Universe is composed of Dark-Matter (DM) particles of unknown nature . 
In order to fit the data, DM particles should be weakly coupled to Standard-Model particles and dynamically relatively cold. 
No direct indication on the mass scale can be inferred from the observations, but the Weakly-Interacting Massive Particle (WIMP), ranging from GeV to few-TeV mass, is a well motivated hypothesis.
All evidences for DM are purely of gravitational origin and, in order to demonstrate that this is indeed the correct interpretation of the observations and to understand its nature, a non-gravitational signal is needed.
To this aim, we can rely on multiple approaches, that well complement each other (e.g. \cite{Fornengo:2016jnx} and reference therein). 

A promising perspective is to detect a DM signal coming from dense DM sites, like our own Galaxy halo. 
In this case, beside direct signal search, a whole and extensive host of astrophysical signals is under intense investigation. 
The idea is that thermal DM relic from the early universe could self annihilate or decay with Standard-Model final states. 
The available observational channels depend on the mass of the DM particle, that annihilate almost at rest, and on the astrophysical background in these channels.
In the WIMP hypothesis, also baryonic and leptonic particles can be created in the halo and a DM signal could be detected looking at the rare CR antimatter components. 
Several interesting hints emerged in the past decade from gamma-rays~\cite{HooperGoodenough:2011}, positrons~\cite{Adriani:2009,Aguilar:2013,Ackermann:2012}  and antiprotons~\cite{CuocoCui:2017}. 
All of these results underline the fact that complementary detection, together with a background-free technique, is highly recommended to shed light on dark matter signatures and candidates.

\begin{figure*}
\begin{center}
  \includegraphics[width=0.38\textwidth]{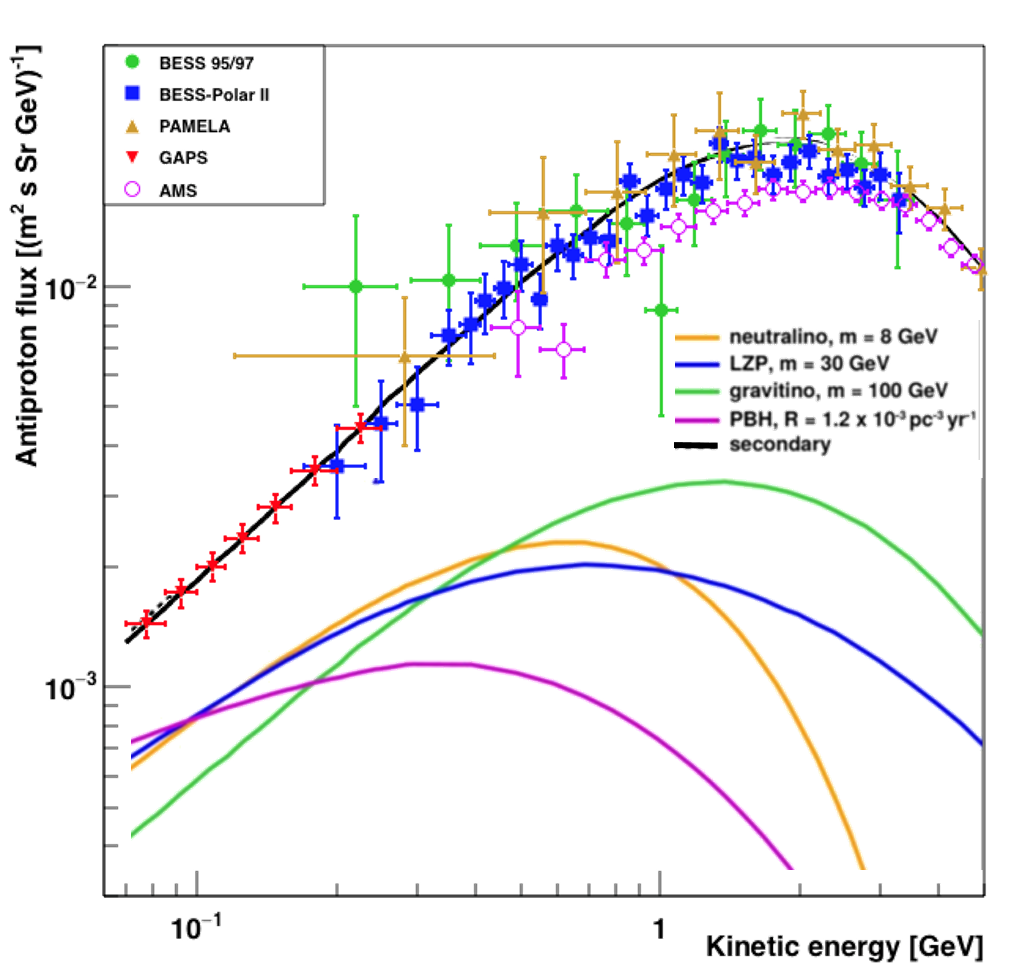}
  \includegraphics[width=0.38\textwidth]{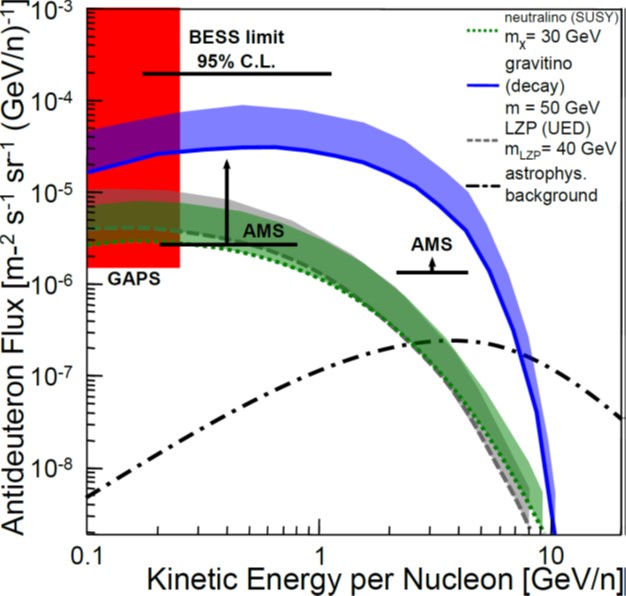}
\end{center}
\caption{
Left: measured $\bar{p}$ flux by BESS\cite{Orito:2000,Abe:2012}, PAMELA\cite{Adriani:2010} and AMS-02\cite{Aguilar:2016}, compared with theoretical calculations for secondary $\bar{p}$ production and for some DM models predicting a sizable $\bar{p}$ contribution at low energy~\cite{Aramaki:2014oda}. 
The expected GAPS sensitivity for one 35-day flight is also shown for the case of secondary production. 
Right: predicted $\bar{d}$ flux as a function of kinetic energy per nucleon by secondary production in the ISM and by some DM particle models. 
The measured limit from BESS is shown, along with the  sensitivities of AMS-02, assuming 5 years of operation, and GAPS, after three 35-day flights~\cite{Aramaki:2016,Aramaki:2013}.  
The arrows on the AMS-02 sensitivities illustrate the geomagnetic-efficiency correction factor calculated along the ISS orbit.
The shaded bands represent the envelope, within medium and maximum effect, of model predictions obtained by considering the uncertainties on the propagation parameters.  
}
\label{fig:1}       
\end{figure*}

Antiprotons ($\bar{p}$) are the most abundant baryonic antiparticle component in CRs. 
Their abundance has been extensively measured by magnetic-spectrometer experiments \cite{Orito:2000,Abe:2012,Adriani:2010,Aguilar:2016} from 200 MeV up to 450 GeV, where it has been found overall consistent, within the uncertainties, with the expected astrophysical background. 
By extending the $\bar{p}$ measurement at lower energies, where the background is kinematically suppressed, it would be possible to probe unexplored DM phase space. 
As illustrated in  Fig.~\ref{fig:1} (left), DM signatures from light neutralinos, gravitinos and Kaluza-Klein particles, as well as evaporating primordial black holes,  could be observed \cite{Aramaki:2014oda}.

Likewise to $\bar{p}$, antinuclei can be produced, both in the ISM and by DM. 
The most favourable signal-to-background ratio is expected for the antideuteron ($\bar{d}$) \cite{Donato:2000,Aramaki:2016}, which production by collisions is more strongly suppressed than by annihilation or decay of massive DM particles, due to the kinematics of the reactions. 
According to several calculations ( Fig.\ref{fig:1} right), below 1~GeV/n  the $\bar{d}$ flux from DM exceeds the background level by more than two orders of magnitude.  

The General Antiparticle Spectrometer (GAPS) experiment is designed to carry out a sensitive DM search by measuring low-energy cosmic-ray baryonic antiparticles, focusing on $\bar{d}$ search.

\section{Low-energy antimatter detection with exotic atoms}
\label{antimatter}

\begin{figure}
\begin{center}
  \includegraphics[width=0.50\textwidth]{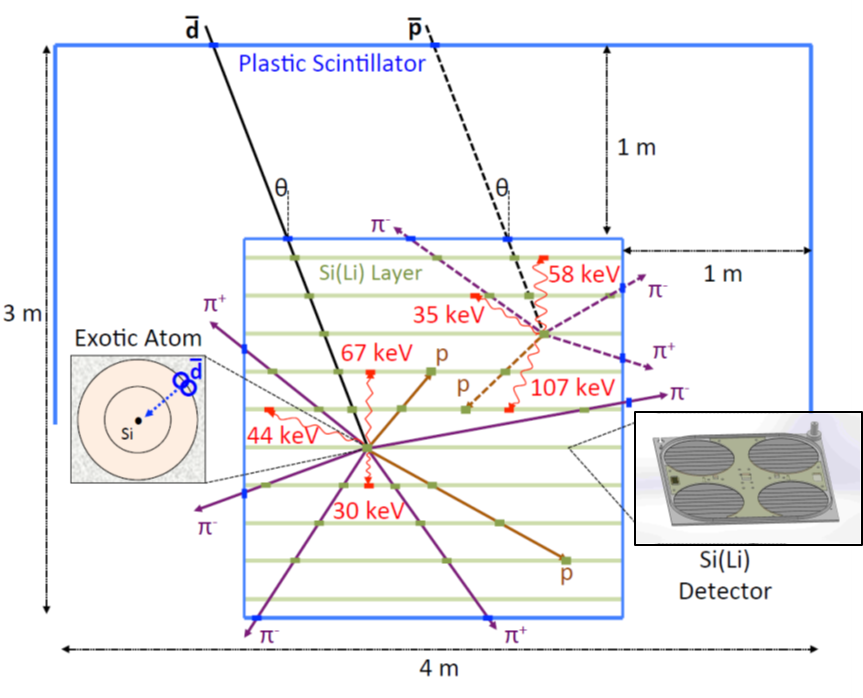}
\end{center}
\caption{GAPS antiparticle detection method: antiparticles slow down and stop in the Si(Li) target, forming an exotic atom. Atomic X-rays will be emitted as it de-excites, followed by the pion (and proton) emission from nuclear annihilation. $\bar{d}$/$\bar{p}$ identification is based on  (1) the stopping range, (2) the pion and proton multiplicity,  (3) the atomic X-rays energies.}
\label{fig:3}       
\end{figure}

The detection principle of the experiment relies on the identification of the antiparticle  annihilation pattern and it is schematically illustrated in  Fig.~\ref{fig:3}. 
The instruments consists of a  time-of-flight system that surrounds a tracking systems. 
Low energy antiparticles slow down in the apparatus until they undergo nuclear annihilation, in flight or at rest, producing a star of pions and  protons. 
Once at rest, before annihilating, they are eventually captured in the medium to form exotic excited atoms, which de-excite through radiative transitions, by emitting detectable X-rays of characteristic energy. 
The simultaneous measurement of the stopping depth, the velocity and the dE/dx loss of the primary antiparticle, of the X-ray energies and of the star particle-multiplicity provides the necessary rejection power to identify the antiparticles, against the particle background, and to distinguish $\bar{d}$ from $\bar{p}$ \cite{Aramaki:2016b}.


An accelerator test was conducted in 2004 and 2005 at KEK, Japan, in order to prove the concept and to precisely measure the X-ray yields of antiprotonic exotic atoms formed with different target materials~\cite{Aramaki:2013}.
An high yield, of about $\sim$75$\%$, was obtained for the low $n$ states, from the de-excitation of antiprotonic exotic atoms with Al and S targets. 
A simple, but comprehensive, cascade model has been developed, aiming to evaluate the X-ray yield for any negatively charged cascading particles with any target materials. 
Three leading de-excitation processes have been considered: Auger, radiative and nuclear-capture transitions. 
The nuclear capture of the cascader might terminate the de-excitation cascade before the exotic atom reaches the ground state. 
The model has been tuned on beam-test data and benchmarked agains other antiprotonic and muonic atoms.  
Hence, the cascade model has been used to evaluate the X-ray yields for antiprotonic and antideuteronic exotic atoms with GAPS materials; e.g. for antideuteronic (antiprotonic) atoms with Si target the predicted yield is $\sim$80$\%$, for X-rays of energy 30, 44 and 67 keV (35, 58 and 106 keV).


\section{ The GAPS instrument design and mission status}
\label{detection}

\begin{figure*}
  \includegraphics[width=1\textwidth]{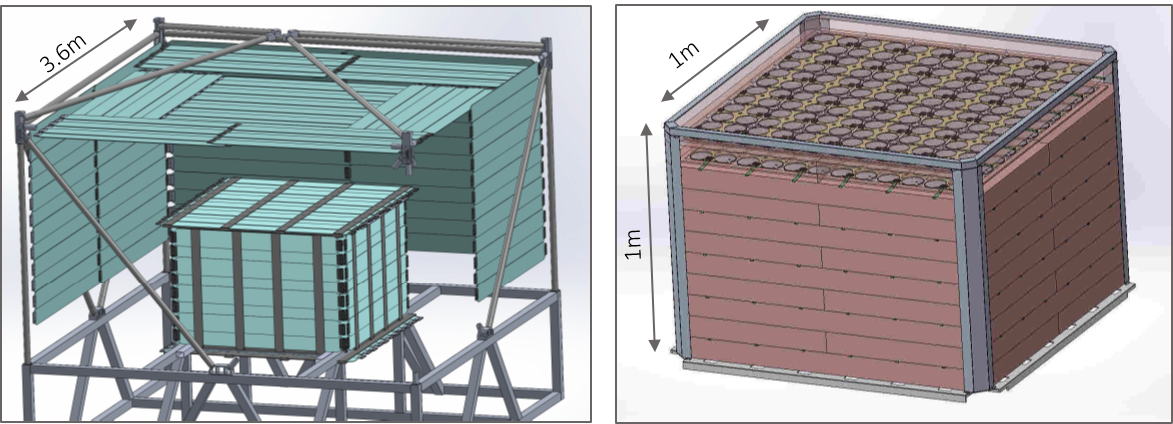}
\caption{GAPS payload design. The outer and inner  plastic-scintillator layers of the TOF system (left) surround a 1~m$^3$ volume, filled with  10 layers of Si(Li) detectors nested into expanded polystyrene blocks (right). Note that not all the TOF outer paddles are  shown in the figure, in order to display the inner layers.}
\label{fig:2}       
\end{figure*}

The detection technique adopted by GAPS does not suffer of the technical limitations of heavy magnets in conventional magnetic spectrometers and allows to easily build a detector with a large geometric acceptance. 

The primary detector is a $\sim$1m$^3$ central volume, containing 10 layers of Lithium-drifted Silicon (Si(Li)) detectors (Fig.~\ref{fig:2} right), which act both as target material for antiparticle annihilation and as tracking device.
Each tracking layer is composed by 12$\times$12 Si(Li) wafers, of 10~cm diameter and 2.5~mm thickness, segmented into 8 strips and mounted on 2$\times$2  Al modules (see the insert in  Fig.~\ref{fig:3}).  
The detector modules are nested into expanded polystyrene blocks, which  provide insulation and protection during chute shock and landing, while minimizing the interference with particle detection. 
The Si(Li) detector performs both as  a 20$\div$80~keV X-ray spectrometer, with 4~keV resolution, and as a tracking system for charged particles, which deposit up to tens of MeVs in the Si(Li) sensors.
The Si(Li) sensors are kept at the operational temperature of -48$^{\circ}$ by means of an Oscillating Heat Pipe (OHP) passive cooling system, developed by JAXA/ISAS for the GAPS experiment \cite{ohp}.

The tracking volume is surrounded by a Time-Of-Flight (TOF) system, which measures the particle velocities and energy depositions, and provides the high-speed trigger.   
The flight distance between the outer and the inner layers is $\sim$1~m.   
The baseline design of the TOF uses thin (5~mm thick) scintillation counters, which have dimensions of 180$\times$16~cm$^{2}$, for the outer TOF layers, and 160$\times$16~cm$^{2}$ for the inner one.   
Approximately 220 scintillation counters are required in total.
Three 6$\times$6~mm$^2$ silicon photomultipliers will be mounted directly on each end of the scintillators.
In order to achieve the needed mass separation, a TOF resolution of 500~ps is required.
An overall illustration of the TOF system is shown in  Fig.~\ref{fig:2} (left).

A prototype flight (pGAPS) was carried out in June 2012  from Taiki, Japan, with the purpose of testing the performance of the GAPS instrument subsystems and the novel OHP thermal cooling concept, as well as measuring the background levels \cite{Mognet,VonDoe}.

The first scientific Long-Duration Balloon (LDB) flight has been recently approved by NASA and has been scheduled for the Antarctic summer of 2020/2021. 
With the full acceptance, in a single flight, GAPS is expected to collect 100 times more statistics of $\bar{p}$, below 250 MeV, than currently available. 
This will  allow to explore unconstrained parameter space for several DM and PBH models (see  Fig.~\ref{fig:1} left). 
Full $\bar{d}$ sensitivity will be reached after $\sim$100~days flight, corresponding to about 3 LDB flights.


\section{ Conclusions}
In the quest to understand dark matter, searching for low-energy antinuclei, in particular antideuterons, is a very promising, but largely unexplored, technique. 
The GAPS experiment, a large-acceptance cosmic-ray balloon instrument, will push the sensitivity limit for antideuterons by two orders of magnitude compared to the best present-day measurements. 
The detection principle of the experiment relies on the identification of the antiparticle  annihilation pattern, including the X-rays emitted from the de-excitation of exotic atoms. 
The development of the science payload of GAPS is now well underway with a potential first flight occurring in late 2020. 
One flight will not be enough to reach the required antideuteron sensitivity, but the instrument is expected to collect 100 times more statistics of low-energy antiprotons than currently available.


\begin{thebibliography}{}
%
%

\bibitem{Fornengo:2016jnx}
  N.~Fornengo,
  arXiv:1701.00119 [astro-ph.HE].

\bibitem{HooperGoodenough:2011} 
D.~Hooper, and L.~Goodenough 
Phys.\ Lett.\  {\bf B697} (2011) 412


\bibitem{Adriani:2009}
O.~Adriani  {\it et al.},
Nature, {\bf 458} (2009) 607

\bibitem{Aguilar:2013}
M.~Aguilar {\it et al.},
Phys.\ Rev.\ Lett.\ {\bf 110 } (2013) 141102

\bibitem{Ackermann:2012}
 M.~Ackermann {\it et al.},
Phys.\ Rev.\ Lett.\ {\bf108} (2012) 011103


\bibitem{CuocoCui:2017} 
M.~Y.~Cui, Q.~Yuan, Y.~L.~S.~Tsai, and Y.~Z.~Fan,
Phys.\ Rev.\  Lett.\, {\bf 118} (2017) 191101;
A.~Cuoco, M.~Krmer, and M.~Korsmeier
Phys.\ Rev.\ Lett.\, {\bf 118} (2017) 191102


\bibitem{Orito:2000}
S.~Orito {\it et al.},
Phys.\ Rev.\ Lett.\ {\bf 84} (2000) 1078–1081

\bibitem{Abe:2012}
K.~Abe {\it et al.},
Phys.\ Rev.\ Lett.\ {\bf 108} (2012) 051102 

\bibitem{Adriani:2010}
O.~Adriani {\it et al.},
Phys.\ Rev.\ Lett.\ {\bf 105} (2010) 121101

\bibitem{Aguilar:2016}
M.~Aguilar {\it et al.},
Phys.\ Rev.\ Lett.\ {\bf 117} (2016) 091103



\bibitem{Donato:2000}
F.~Donato, 
N.~Fornengo and 
P.~Salati
Phys.\ Rev.\ D {\bf 62} (2000) 043003


\bibitem{Aramaki:2016}
  T.~Aramaki {\it et al.},
  Phys.\ Rep.\  {\bf 618} (2016) 1
  
\bibitem{Aramaki:2013}
  T.~Aramaki {\it et al.},
  Astropart.\ Phys.\   {\bf 49} (2013) 52, 

\bibitem{Aramaki:2014oda}
  T.~Aramaki {\it et al.} [GAPS Collaboration],
  Astropart.\ Phys.\  {\bf 59} (2014) 12

\bibitem{Aramaki:2016b}
  T.~Aramaki {\it et al.},
  Astrp.\ Ph.\  {\bf 74} (2016) 6


\bibitem{ohp}
S.~Okazaki {\it et al.} 
 J.\ Astr.\ {\bf 3} (2014) 


\bibitem{Mognet}
S.~A.~I.~Mognet {\it et al.} 
Nucl.\ Instr.\ Meth.\ {\bf 735} (2014)

\bibitem{VonDoe}
P.~Von Doetinchem  {\it et al.} 
Astro.\ Ph.\ {\bf 54} (2014) 93 



\end{thebibliography}
\end{document}